\documentclass[applemac]{aa}
\usepackage{inputenc}
\usepackage{natbib}
\usepackage{txfonts}
\usepackage{graphicx}
\usepackage{color}
\bibpunct{(}{)}{;}{a}{}{,}

\begin{document}

\title{The effect of multilayer ice chemistry on gas-phase deuteration in starless cores}
\author{O. Sipil\"a\inst{1},
		P. Caselli\inst{1},
	     \and{V. Taquet\inst{2}}
}
\institute{Max-Planck-Institute for Extraterrestrial Physics (MPE), Giessenbachstr. 1, 85748 Garching, Germany \\
e-mail: \texttt{osipila@mpe.mpg.de}
\and{Leiden Observatory, Leiden University, P.O. Box 9513, 2300-RA Leiden, The Netherlands}
}

\date{Received / Accepted}

\abstract
{Astrochemical models commonly used to study the deuterium chemistry in starless cores consider a two-phase approach in which the ice on the dust grains is assumed to be entirely reactive. Recent experimental studies suggest that cold interstellar ices are mostly inert, and a multilayer model distinguishing the chemical processes at the surface and in the ice bulk would be more appropriate.}
{We aim to investigate whether the multilayer model can be as successful as the bulk model in reproducing the observed abundances of various deuterated gas-phase species toward starless cores.}
{We calculate abundances for various deuterated species as functions of time using a pseudo-time-dependent chemical model adopting fixed physical conditions. We also estimate abundance gradients in starless cores by adopting a modified Bonnor-Ebert sphere as a core model. In the multilayer ice scenario, we consider desorption from one or several monolayers on the surface.}
{We find that the multilayer model predicts abundances of $\rm DCO^+$ and $\rm N_2D^+$ that are about an order of magnitude lower than observed, caused by the trapping of CO and $\rm N_2$ into the grain mantle. As a result of the mantle trapping, deuteration efficiency in the gas phase increases and we find stronger deuterium fractionation in ammonia than what has been observed. Another distinguishing feature of the multilayer model is that $\rm D_3^+$ becomes the main deuterated ion at high density. The bulk ice model is generally easily reconciled with observations.}
{Our results underline that more theoretical and experimental work is needed to understand the composition and morphology of interstellar ices, and the desorption processes that can act on them. With the current constraints, the bulk ice model appears to be better in reproducing observations than the multilayer ice model. According to our results, the $\rm H_2D^+$ to $\rm N_2D^+$ abundance ratio is higher than 100 in the multilayer model, while only a few $\times~10$ in the bulk model, and so observations of this ratio could provide information on the ice morphology in starless cores. Observations of the abundance of $\rm D_3^+$ compared to $\rm H_2D^+$ and $\rm D_2H^+$, although challenging, would provide additional constraints for the models.}


\maketitle

\section{Introduction}

Deuterated species are useful tracers of the cold ($T \sim 10\, \rm K$) and dense ($n_{\rm H} \sim 10^4 - 10^6 \, \rm cm^{-3}$) interiors of starless cores, where otherwise abundant neutral species, such as CO, are depleted onto grain surfaces. After depletion occurs, deuterium chemistry initiates through the exothermic reactions
\begin{eqnarray}
\rm H_3^+ + HD &\longrightarrow& \rm H_2D^+ + H_2 \\
\rm H_2D^+ + HD &\longrightarrow& \rm D_2H^+ + H_2 \\
\rm D_2H^+ + HD &\longrightarrow& \rm D_3^+ + H_2 \, .
\end{eqnarray}
The deuterated $\rm H_3^+$ isotopologs can then transfer a deuteron to other species (for example, CO or $\rm N_2$), forming various deuterated ions, and also contribute to the formation of deuterated ammonia \citep{Rodgers01}. As a result of the deuteration process, the deuterium fraction in the various species, i.e., the abundance ratio of the deuterated and hydrogenated isotopologs, can increase to orders of magnitude above the elemental D/H ratio of $\sim$$10^{-5}$ \citep{Linsky03}. Indeed, modeled deuterium fractions can easily reach $\geq 10\,\%$ \citep{Roberts03}, and such high fractionation has also been observed for example for ammonia \citep{Roueff05}, formaldehyde \citep{Bergman11}, and methanol \citep{Parise04}.

The dissociative recombination of deuterated ions releases atomic deuterium into the gas, which can then be adsorbed onto grain surfaces and thus contributes to deuteration on the surface by addition or abstraction reactions. The efficiency of surface reactions is influenced heavily by the properties of the ice \citep[e.g.][]{Taquet12}. Many past models of surface chemistry have adopted a so-called bulk ice approach where the entirety of the ice on a grain is available for chemical reactions and (thermal or non-thermal) desorption \citep[recently:][]{Semenov10, Aikawa12, Sipila13}. However, interstellar ices are expected to be amorphous, layered structures \citep{Hama13}, and the bulk approach seems unrealistic as a model for the grain chemistry given the experimental evidence against efficient reactivity beneath a few monolayers (MLs) of ice \citep{Watanabe03, Watanabe04}. As an alternative to the bulk model, three-phase models separating the ice into a reactive surface layer and an unreactive mantle have been studied \citep{HH93b, Charnley09, Garrod11, Taquet12}. The three-phase model is here referred to as the multilayer model. The multilayer approach has advantages over the bulk approach, in which every atom/molecule can in principle react with any other atom/molecule in the ice. The multilayer model allows for a more intuitive description of the overall reactivity by limiting the reactions to a part of the ice, although everything can still react with everything within that part of the ice. The multilayer approach therefore affects the chemical composition of ices by trapping reactive species into the bulk, inducing higher abundances of radicals in the ice in the multilayer model compared to the bulk model \citep{Taquet12}. Another example of the advantages of the multilayer approach is the treatment of desorption; in the bulk model species can be desorbed from anywhere in the ice matrix whereas in the (conventional) multilayer model desorption only occurs from the top layer(s), which is physically more reasonable.

The previous multilayer models have mostly concentrated on studying the evolution of ice species. In this paper, we want to concentrate instead on the feedback of multilayer ice chemistry to the chemical evolution of deuterated species in the gas phase, with the particular aim of investigating whether the multilayer ice model can reproduce the observed abundances of various deuterated species as successfully as the bulk model has been found to do. This is accomplished by updating our previous gas-grain models \citep{Sipila15a, Sipila15b}, which include extensive descriptions of deuterium and spin-state chemistry, with the multilayer ice approach presented by \citet{HH93b}. The abundances of the various species are studied in conditions with varying density and temperature, and comparisons between the bulk and multilayer approaches in predicted gas-phase abundances are made.

The paper is structured as follows. In Sect.\,2, we present the chemical model used in this paper. In Sect.\,3, we present results from single-point models, assuming fixed values for the physical parameters, and from a core model yielding abundance gradients. We discuss our results and present some alternative models in Sect.\,4, and present our conclusions in Sect.\,5. Appendix~A presents some additional results.

\section{Chemical model}

\subsection{Multilayer and bulk ice chemistry}\label{ss:modeldesc}

In this paper, we use a modified version of the gas-grain chemical code described in detail in \citet{Sipila15a}. Specifically, we have added to the code the three-phase description of gas-grain chemistry developed by \citeauthor{HH93b}~(\citeyear{HH93b}; see also \citealt{Taquet14}), in which the ice on the grains is separated into a chemically active surface layer and a chemically inert mantle. Chemical species can be transferred from the surface to the mantle in the case of net adsorption onto the grain surfaces, and vice versa in the case of net desorption. As noted in the Introduction, this approach allows for a (qualitatively) more realistic approach to grain-surface chemistry compared to bulk ice models (see \citealt{Sipila12}, \citealt{Sipila13} and \citealt{Sipila15a} for details of our previous bulk ice model). In this paper, we compare a bulk ice model to a multilayer model to assess the possible impact of a different ice description on the gas-phase abundances of deuterated species.

The present model includes three desorption processes. Thermal desorption and cosmic ray (CR) induced desorption \citep{HH93} are treated identically to \citet{Sipila15a}. In addition, we have introduced CO and $\rm H_2O$ photodesorption into the model using the formulae presented in \citeauthor{Coutens14}~(\citeyear{Coutens14}; see also \citealt{Hollenbach09}). The assumed photodesorption yields are $2.7\times10^{-3}$ for CO \citep{Oberg09a} and $1.0\times10^{-3}$ for $\rm H_2O$ \citep{Oberg09b}. For simplicity, we assume that HDO and $\rm D_2O$ photodesorb with the same yield as $\rm H_2O$, and that this applies also to the associated ortho and para states. We assume that $\rm H_2O$ photodesorbs in molecular form, although studies have shown that its dissociation products can desorb individually upon absorption of a UV photon by a surface $\rm H_2O$ molecule \citep{Andersson08, Arasa10, Arasa15}. We assume $F_0 = 10^8 \rm \, photons \, cm^{-2} \, s^{-1}$ for the flux of interstellar photons and $G_{\rm CR} = 10^{-4}$ for the efficiency of photodesorption by secondary UV photons created by $\rm H_2$ excitation in environments with high visual extinction \citep{Prasad83, Shen04}. In the bulk ice model no restrictions are placed on thermal and cosmic ray induced desorption, as in our previous studies. However, photodesorption is limited to two MLs (as also assumed recently by \citealt{Hincelin15}).

In their multilayer model, \citet{HH93b} assumed that desorption processes affect the surface layer only. However, studies have shown that photodesorption can occur from multiple layers, with the photodesorption yield dropping deeper into the ice \citep{Oberg09b, Arasa15}. Furthermore, in the case of CO, \citet{vHemert15} have found that photodesorption inwards from the third monolayer is almost negligible. Since cosmic rays are highly energetic and can easily penetrate deep into the ice, it seems reasonable that cosmic-ray induced desorption could also occur from beneath the surface layer\footnote{A detailed treatment of direct cosmic-ray impacts on icy mantles is beyond the scope of this paper, but is being investigated by Vasyunin et al. following \citet{Ivlev15}.}. We therefore assume that, in the multilayer model, the included desorption processes apply to the surface layer and to one layer in the mantle. (The photodesorption yield is assumed to be equal for surface and mantle photodesorption.)

In practice, mantle desorption is implemented into the model by applying an efficiency factor to the desorption terms of the mantle species. The efficiency is unity until the first monolayer (ML) of the mantle is complete, and drops as 1/ML afterwards; when $n$ MLs of ice ($n \geq 1$) have accumulated in the mantle, the CR desorption terms of the mantle species are multiplied by $1/n$ which ensures that desorption comes from one ML. Essentially, the mantle is treated as a bulk and the desorption rate of each mantle species is dependent on the average abundance of that species in the mantle. The desorption efficiency is unity at all times for species in the surface layer.

\begin{figure*}
\centering
\includegraphics[width=2.0\columnwidth]{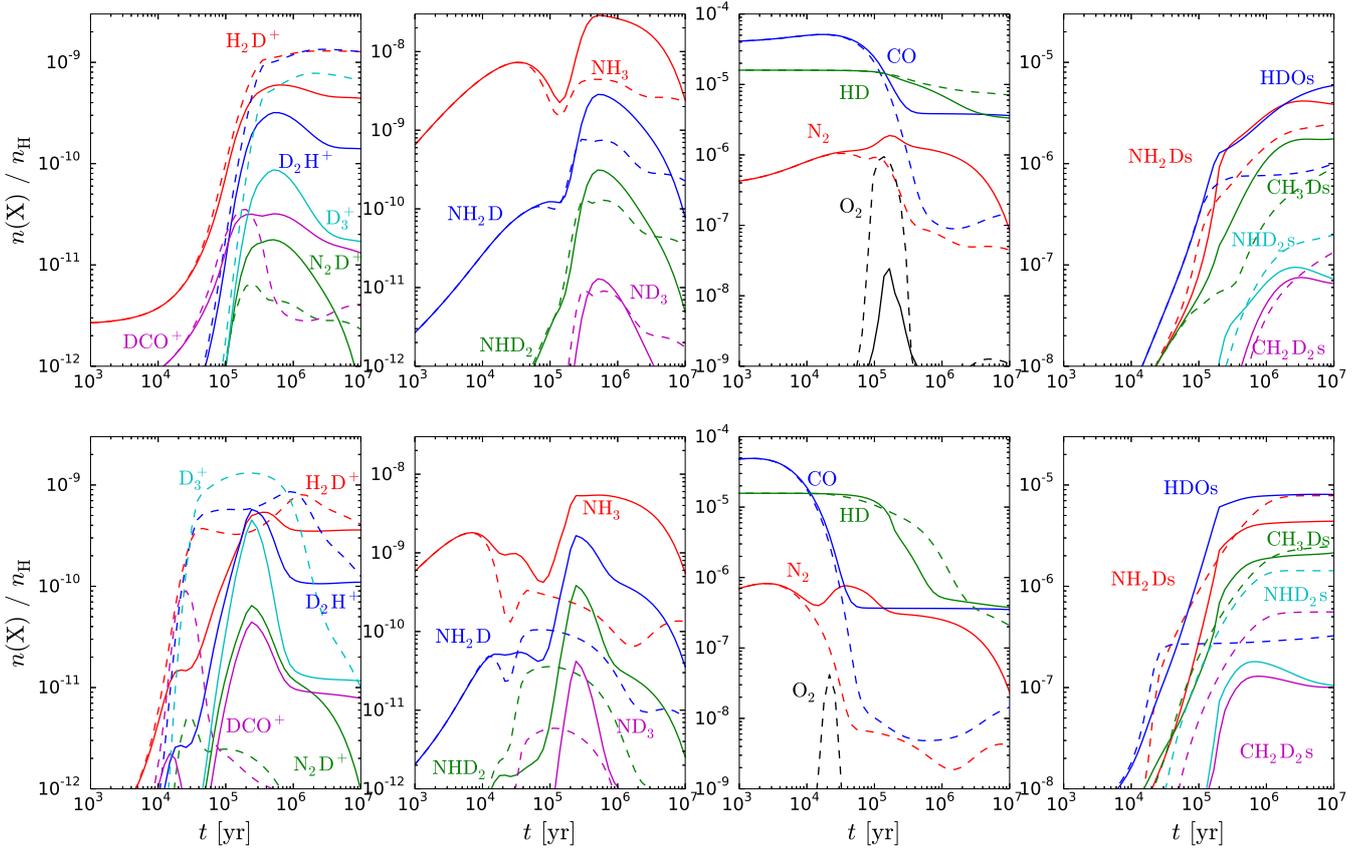}
\caption{Fractional abundances (with respect to $n_{\rm H}$) of selected gas-phase and ice species (labeled in the plots) as functions of time in a homogeneous model with $T_{\rm gas} = T_{\rm dust} = 10$\,K. For those species that have spin states, the plotted abundances represent sums over the spin states. Solid lines correspond to the bulk ice model, while dashed lines correspond to the multilayer ice model. In the latter case, the ice abundances (right-hand panels) represent sums over the surface and mantle. {\sl Upper row}: $n_{\rm H} = 10^5 \rm \, cm^{-3}$. {\sl Lower row}: $n_{\rm H} = 10^6 \rm \, cm^{-3}$.
}
\label{fig:bulk_vs_multi}
\end{figure*}

\subsection{Chemical reaction sets and model parameters}\label{ss:chemsets}

\begin{table}
\caption{Adopted values for the various physical parameters (left column) and adopted initial abundances (right column).}
\centering
\begin{tabular}{c c | c c}
\hline \hline 
Parameter & Value & Species & Initial abundance \\ \hline
$T_{\rm gas} = T_{\rm dust}$ & 10\,K & $\rm H_2$ & 0.5 \\
$\zeta$ & $1.3\times10^{-17} \, \rm s^{-1}$ & $\rm He$ & $9.00\times10^{-2}$  \\
$A_{\rm V}$ & 10 mag & $\rm HD$ & $1.60\times10^{-5}$ \\
$a_{\rm g}$ & $0.1 \, \rm \mu m$ & $\rm O$ & $2.56\times10^{-4}$ \\ 
$\rho_{\rm g}$ & $3.0 \rm \, g \, cm^{-3}$ & $\rm C^+$ & $1.20\times10^{-4}$  \\
$n_s$ & $1.5 \times 10^{15} \, \rm cm^{-2}$ & $\rm N$ & $7.60\times10^{-5}$ \\
$E_{\rm d} / E_{\rm b}$ & 0.77 & $\rm S^+$ & $8.00\times10^{-8}$ \\
$R_{\rm d}$ & 0.01 & $\rm Si^+$ & $8.00\times10^{-9}$ \\
 & & $\rm Na^+$ & $2.00\times10^{-9}$ \\
 & & $\rm Mg^+$ & $7.00\times10^{-9}$ \\
 & & $\rm Fe^+$ & $3.00\times10^{-9}$ \\
 & & $\rm P^+$ & $2.00\times10^{-10}$ \\
 & & $\rm Cl^+$ & $1.00\times10^{-9}$ \\
  & & $\rm H_2\,(o/p)_{\rm ini}$ & $1.00\times10^{-3}$ \\
\hline
\end{tabular}
\tablefoot{The data is reproduced from \citet{Sipila15a}. The parameters presented in the left column represent the (gas and dust) temperature ($T$), cosmic ray ionization rate ($\zeta$), visual extinction ($A_{\rm V}$), grain radius ($a_{\rm g}$), grain material density ($\rho_{\rm g}$), density of binding sites on the grain surface ($n_s$), diffusion energy to binding energy ratio ($E_{\rm d} / E_{\rm b}$), and dust-to-gas mass ratio ($R_{\rm d}$).
}
\label{tab1}
\end{table}

We adopt the gas-phase and grain-surface chemical reaction sets from \citet{Sipila15b}. Both sets include deuterated species with up to six atoms. In \citet{Sipila15b}, the spin-state branching ratios in reactions involving multiple protons and/or deuterons were derived using a group-theoretical approach. As in \citet{Sipila15b}, the present model considers explicitly the spin states of all species included in the $\rm H_3^+ + H_2$ reacting system and in the water and ammonia formation networks, although in what follows we only present results where the spin states of each species have been summed over. In total, the model contains $\sim$$51000$ gas-phase reactions, $\sim$$2600$ grain-surface reactions and $\sim$$1400$ chemical species (the various spin states are counted as distinct species).

In this paper, we present the abundances of various (deuterated) species as functions of time as calculated with our pseudo-time-dependent rate-equation chemical code. We show the results of single-point models (corresponding to fixed values of the physical parameters), and from core models including density and temperature gradients. Table~\ref{tab1} shows the adopted physical parameters and initial chemical abundances, which are taken from \citet{Sipila15a}. These values were used to produce the results presented below, except where otherwise noted.

The grain model adopted here (single-sized spherical grains; parameters given in Table~\ref{tab1}) is comparable to that used in the GRAINOBLE multilayer model \citep{Taquet12, Taquet13, Taquet14}. However, some differences between the models exist. Here we adopt constant binding energies for all species while in the GRAINOBLE model the binding energies vary dynamically with the coverage of the ice \citep{Taquet14}. The GRAINOBLE model assumes desorption from the surface layer only while we allow desorption from one layer in the mantle as well. We compare the present chemical model to the GRAINOBLE model in Sect.\,\ref{ss:taq_comparison}.

All of the other details of the chemical model, including the forms of the rate coefficients for the various chemical processes considered, can be found in \citet{Sipila15a}.

\section{Results}

\subsection{Single-point models}

\subsubsection{Variable density}\label{sss:vartemp}

Figure~\ref{fig:bulk_vs_multi} shows the abundances of some common (deuterated) gas-phase species along with the most abundant deuterated ice constituents as functions of time for the bulk ice and multilayer models and different physical conditions. It is evident that the results can be very different depending on the choice of the ice model.

At $n_{\rm H} = 10^5 \rm \, cm^{-3}$, the abundances predicted by the two models agree well up to $t \sim 10^5\,\rm yr$, but at later times the multilayer model predicts a factor of $\sim$2 more $\rm H_2D^+$, a factor of $\sim$10 more $\rm D_2H^+$, and a factor of $\sim$40 more $\rm D_3^+$ (at late times), while simultaneously the abundances of $\rm N_2D^+$ and $\rm DCO^+$ can be a factor of $\sim$3-4 and 10 lower, respectively, than in the bulk model. This can be understood as follows. $\rm H_2D^+$, $\rm D_2H^+$, and $\rm D_3^+$ are most efficiently destroyed by CO for as long as it subsists in the gas phase. Mantle trapping leads to very efficient CO freezeout and consequently the abundances of deuterated $\rm H_3^+$ are able to attain higher values in the multilayer model than in the bulk model. Similarly, the abundances of $\rm N_2D^+$ and $\rm DCO^+$ suffer from the mantle trapping of their precursor molecules.

The $\rm H_3^+$ isotopologs are formed mainly by reactions belonging to the $\rm H_3^+ + H_2$ reacting system \citep[e.g.,][]{Hugo09} and are thus dependent on the abundance of gas-phase HD. In the bulk model, the deuteration degree begins to decrease at long timescales caused by the depletion of HD from the gas phase by grain-surface chemistry \citep[see also][]{Sipila13}, whereas in the multilayer model H, $\rm H_2$, and their deuterated isotopologs have a very limited amount of reacting partners which leads to boosted surface HD production through $\rm H + D \longrightarrow HD$, retaining the degree of deuterium fractionation in the gas phase.

Ammonia presents an interesting behavior, as the differences between the bulk and multilayer models are smaller the more D atoms are involved. Ammonia production starts with the dissociation of $\rm N_2$ by $\rm He^+$, and is thus dependent on the depletion of nitrogen onto grain surfaces. The depletion is stronger in the multilayer model because of mantle trapping. However, ammonia deuteration is controlled mainly by the $\rm H_3^+$ isotopologs \citep[see][]{Rodgers01, Sipila15b}, and thus the increased abundances of the $\rm H_3^+$ isotopologs in the multilayer model compensate for the depletion and act to increase the deuteration degree of ammonia with respect to the bulk model.

\begin{figure*}
\centering
\includegraphics[width=1.85\columnwidth]{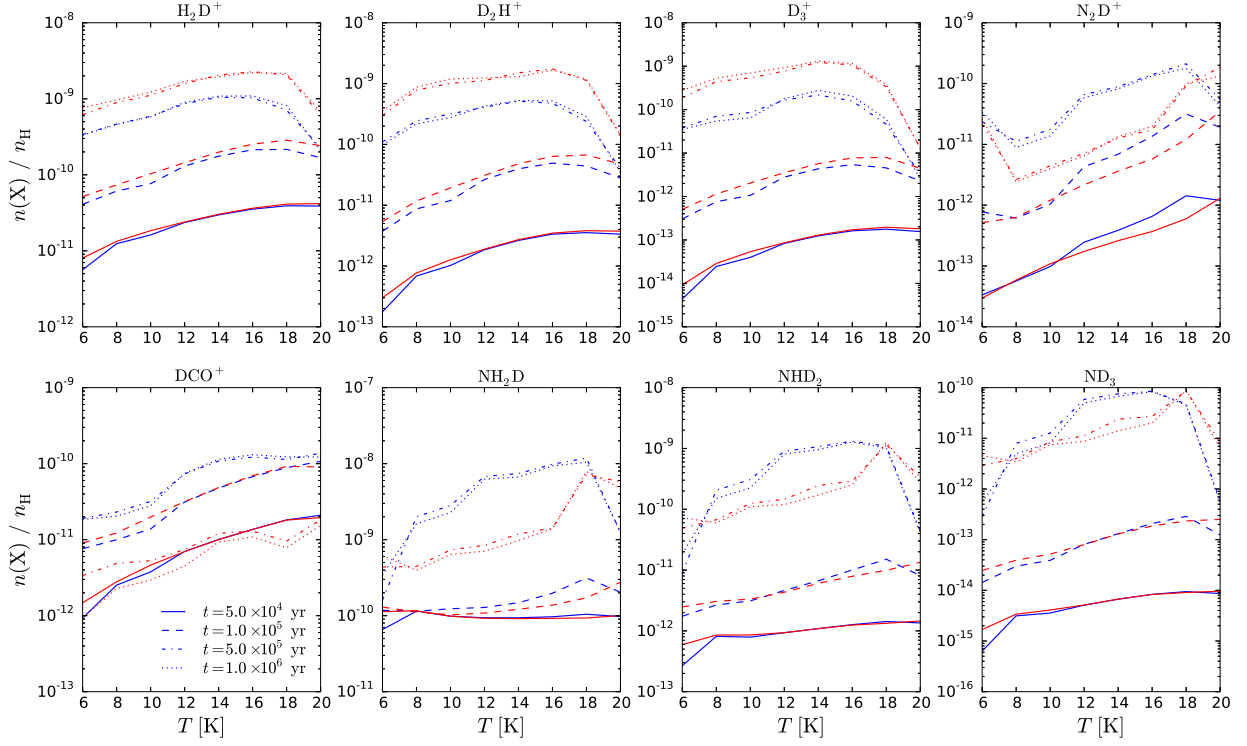}
\caption{Abundances of various deuterated species as functions of temperature ($T_{\rm gas} = T_{\rm dust}$) at different times (indicated in the figure). The adopted medium density is $n_{\rm H} = 10^5 \rm \, cm^{-3}$. The blue lines represent the bulk ice model, while the red lines represent the multilayer ice model.
}
\label{fig:bulk_vs_multi_temp_nh1e5}
\end{figure*}

\begin{figure*}
\centering
\includegraphics[width=1.85\columnwidth]{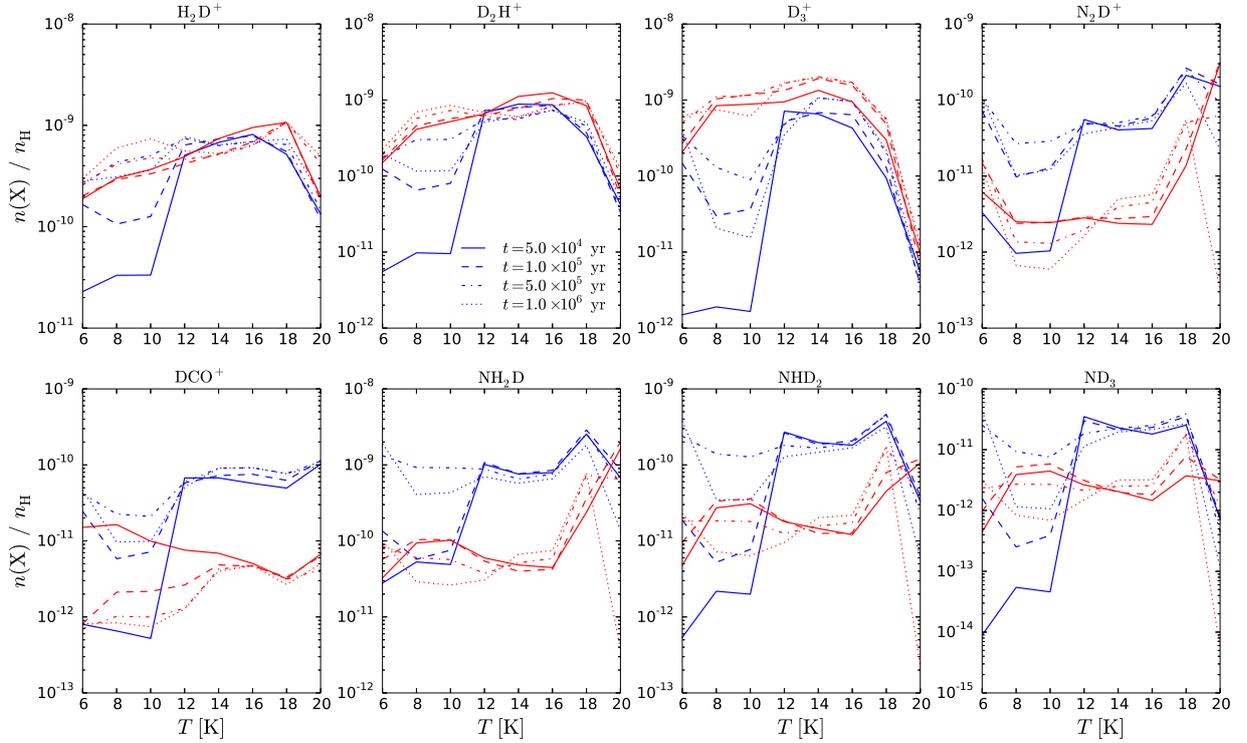}
\caption{As for Fig.\,\ref{fig:bulk_vs_multi_temp_nh1e5}, but for $n_{\rm H} = 10^6 \rm \, cm^{-3}$.
}
\label{fig:bulk_vs_multi_temp_nh1e6}
\end{figure*}

In both models, the three main deuterium carriers on the grain surface are HDO, $\rm NH_2D$ and $\rm CH_3D$. As explained above, HD production is favored on grains in the multilayer model because of the lack of atomic deuterium transfer to the mantle, leading to decreased deuterium fractionation in the ice in the multilayer model as compared to the bulk model. We note that unlike in \citet{Sipila13}, deuterium is distributed more evenly even in the bulk model, as opposed to getting locked mostly in surface $\rm HDO$. This is partly due to a missing dissociation reaction ($\rm HDO \longrightarrow OH + D$ on the surface) in the \citet{Sipila13} model which was added for Sipil\"a et al.\,(\citeyear{Sipila15a}; see their Sect.\,3.1), but mainly because of the introduction of deuteration to up to six atoms which allows for complete treatments of ammonia and methane deuteration (which depend critically on $\rm NH_4^+$ and $\rm CH_5^+$, respectively).

{At $n_{\rm H} = 10^6 \rm \, cm^{-3}$, the results of the bulk and multilayer models are very different. Considering first the multilayer model only, it is seen that the deuteration peak shifts from a few $\times 10^5$\,yr to very early times, a few $\times 10^4$\,yr. This is because the efficiency of mantle transport depends on the total formation rates of the various species on the grain surface \citep{HH93b}; at high density, adsorption rates are high, decreasing the time required to accumulate one ML of ice on the surface. Accordingly, CO and $\rm N_2$ deplete from the gas very fast, leading to extremely low abundances of $\rm DCO^+$ and $\rm N_2D^+$ already before $10^5$\,yr of chemical evolution. Deuterium fractionation is very strong at this density, and $\rm D_3^+$ becomes the most abundant $\rm H_3^+$ isotopolog, an effect which has previously been seen in some gas-phase chemical models \citep{WFP04, FPW04, Pagani09, Sipila10}. Notably, the deuterium fraction in ammonia is much larger than what has been observed toward starless/prestellar cores. This issue is discussed in more detail in Sect.\,\ref{ss:obsimp}. The high deuterium fractionation in the gas is reflected also in the ice, where non-negligible abundances of doubly-deuterated ammonia and methane are found at long timescales. Notably, HDO ice is not very abundant in the multilayer model. This is because of the adopted binding energy of 1390\,K for atomic oxygen \citep{Bergeron08, Cazaux11, Sipila12}\footnote{We note that \citet{He15} have recently presented even higher values for the O binding energy ($1660\pm60\,\rm K$ on porous water ice).} which makes it more difficult to remove atomic O from the mantle, through mantle desorption, once it is transfered there\footnote{$\rm H_2O$ ice is very abundant also in the multilayer model because of the greater mobility of atomic H over atomic D.}. Indeed, for an O binding energy of 800\,K a lot more HDO ice is formed -- we discuss models with a lower oxygen binding energy in Appendix~\ref{aa:oxygen}.

Turning then to the bulk model, it is seen that the abundances of the $\rm H_3^+$ isotopologs begin to decrease at around $10^6$\,yr owing to HD depletion, and $\rm D_3^+$ never becomes the main deuterated ion although its peak abundance is comparable to those of $\rm H_2D^+$ and $\rm D_2H^+$. In contrast to the multilayer model, $\rm N_2D^+$ and $\rm DCO^+$ abundances peak at later times  because of the slower formation of the deuterated $\rm H_3^+$ isotopologs, and appreciable abundances of $\rm N_2D^+$ and $\rm DCO^+$ are retained in the gas even after the deuterium peak. HDO is the main deuterium carrier on the surface (in line with the results of \citealt{Sipila13}), and the deuterium fraction in gas-phase ammonia is lower than in the multilayer model.

The high gas-phase $\rm O_2$ abundance seen around $t \sim 10^5$\,yr at $n_{\rm H} = 10^5 \rm \, cm^{-3}$ in the multilayer model is, somewhat counterintuitively, caused by mantle trapping. In the present model, the binding energy of atomic oxygen is larger than those of N and C (800\,K). Mantle trapping effectively decreases the depletion timescale compared to the bulk model, and the effect is greater for N and C than for O because of their lower binding energies (O is trapped relatively efficiently on the surface even in the bulk model). The faster depletion of N helps OH to form $\rm O_2$ through $\rm O + OH \longrightarrow O_2 + H$ in the gas phase, instead of reacting with N to form $\rm NO + H$, which is the main destruction pathway of OH at $t \sim 10^5$\,yr in the bulk model. $\rm O_2$ desorption from the mantle also contributes to the gas-phase $\rm O_2$ abundance in the multilayer model. However, it is not the sole reason for the high gas-phase $\rm O_2$ abundance as evidenced by the high abundance found also in a model excluding mantle desorption (Sect.\,\ref{ss:nomantledes}; Fig.\,\ref{fig:multi_vs_multi_nomantledes}). The general impact of mantle desorption is further discussed in Sect.\,\ref{ss:nomantledes}, where we show the results of calculations that exclude this process. Finally, we note that the peak $\rm O_2$ abundance predicted by the multilayer model is about a factor of 10-100 higher than what has been deduced from ({\sl SWAS, Odin, Herschel}) observations toward low-temperature objects in the ISM \citep{Goldsmith00, Pagani03, Larsson07, Liseau12, Yildiz13}. In light of the observations the bulk model seems to give more reasonable results, which is also the case for the various deuterated species shown in Fig.\,\ref{fig:bulk_vs_multi}. We return to this issue in Sect.\,\ref{ss:obsimp}.

\subsubsection{Variable temperature}

Figure~\ref{fig:bulk_vs_multi_temp_nh1e5} shows the abundances of selected species as functions of temperature at $n_{\rm H} = 10^5 \rm \, cm^{-3}$ using different ice models. The same trends that were apparent at 10\,K in Fig.\,\ref{fig:bulk_vs_multi} are present at different temperatures as well. On the one hand, multilayer chemistry enhances the abundances of deuterated $\rm H_3^+$ at late times, while on the other hand the abundances of species that contain heavier elements are lower than in the bulk model. However, the two models predict very similar abundances up to $t = 10^5$\,yr. The general deuteration efficiency increases with temperature, peaking at around 15-18\,K depending on the species. A similar trend was found in \citet{Sipila15b}, although in that paper a sparser temperature grid was used. Deuterated ammonia depletes very efficiently at 20\,K on the one hand because of the destruction of the $\rm H_3^+$ isotopologs due to the temperature, and on the other hand because of the depletion of ammonia (see \citealt{Sipila15a}).

Figure~\ref{fig:bulk_vs_multi_temp_nh1e6} shows the results of calculations otherwise similar to Fig.\,\ref{fig:bulk_vs_multi_temp_nh1e5}, but adopting $n_{\rm H} = 10^6 \rm \, cm^{-3}$. Here, the effect of the temperature is more prominent. The multilayer model predicts clearly higher abundances than the bulk model for the $\rm H_3^+$ isotopologs below 12\,K, but the results of the two models are similar at higher temperatures. Notably, the abundance of $\rm D_3^+$ is always higher in the multilayer model. At short timescales, $\rm DCO^+$, $\rm N_2D^+$ and deuterated ammonia are generally more abundant in the multilayer model than in the bulk model at $T < 12\,\rm K$ (see also Fig.\,\ref{fig:bulk_vs_multi}), while the inverse is true at higher temperatures. At late times however the multilayer model presents (much) lower abundances for these species in the entire temperature range.

\subsection{Core models}

To investigate the effect of multilayer chemistry on chemical abundance profiles in starless core conditions, we calculated radial profiles for the abundances of various species using a modified Bonnor-Ebert sphere \citep{Evans01, Keto05, Sipila11, Sipila15c} as the core model. For this study, we chose a model core with a high non-dimensional radius ($\xi = 16$) in order to get a density gradient of about two orders of magnitude across the core, noting that this configuration is highly supercritical and thus unstable against gravitational contraction \citep{Sipila15c}. We assumed $M = 1\,M_{\odot}$ for the core mass and $A_{\rm V} = 2$\,mag for the external visual extinction.

Following the approach discussed in our previous papers (see for example \citealt{Sipila15b}), we divided the core model into spherical shells and calculated the chemical evolution separately in each shell. The initial chemical abundances, given in Table~\ref{tab1}, are the same in each shell. The gas temperature is expected to change as a function of time because the depletion of the coolant species (mainly CO) decreases the cooling rates at low densities (up to a few $\times~10^4 \, \rm cm^{-3}$) where the gas and dust are not efficiently coupled \citep{Goldsmith01, Juvela11}, and so cores at various stages of development should display different gas temperature profiles. We modeled this effect by extracting the abundances of various cooling species from the initial core model (which assumes $T_{\rm dust} = T_{\rm gas}$) at two timesteps, $t = 1 \times 10^5 \, \rm yr$ and $t = 5 \times 10^6 \, \rm yr$, and using the profiles to calculate the gas temperature (see, e.g., \citet{Sipila12} for a detailed explanation of this process). Figure~\ref{fig:DT} presents the resulting density and temperature structures. In what follows, we label the density and temperature structures corresponding to $t = 1 \times 10^5 \, \rm yr$ and $t = 5 \times 10^6 \, \rm yr$ as DT1 and DT2, respectively.

As discussed in \citet{Sipila15c}, these results do not (necessarily) represent the development of a core from one evolutionary state to the next, but show two possible structures for an $M = 1\,M_{\odot}$, $\xi = 16$ core in environments with different chemical ages. Also, comparison with Fig.~3 in \citet{Sipila15c} shows that the gas temperature profiles calculated in this paper are lower for the (chemically) younger core. The main reason for this is that in the present paper we do not include quantum tunneling through activation barriers in grain-surface reactions, which helps to maintain a higher CO abundance in the ice -- and in the gas phase through desorption -- and hence more efficient gas cooling than in the \citet{Sipila15c} paper, where the chemical model of \citet{Sipila13} was adopted.

\begin{figure}
\centering
\includegraphics[width=1.0\columnwidth]{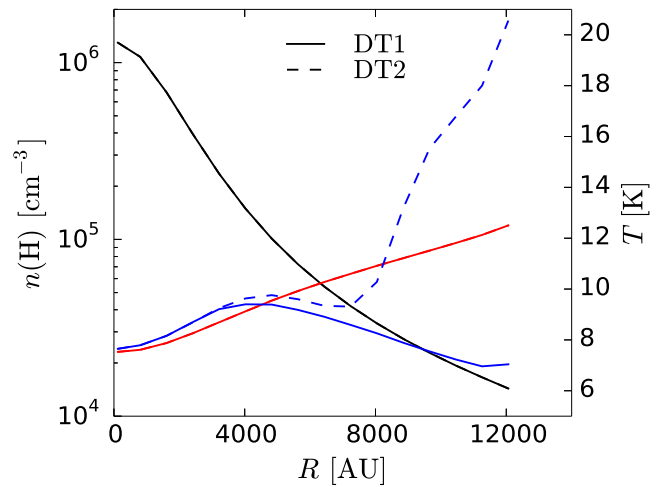}
\caption{Density (black), dust temperature (red), and gas temperature (blue) profiles of the core models DT1 (solid lines) and DT2 (dashed lines).
}
\label{fig:DT}
\end{figure}

\begin{figure*}
\centering
\includegraphics[width=1.9\columnwidth]{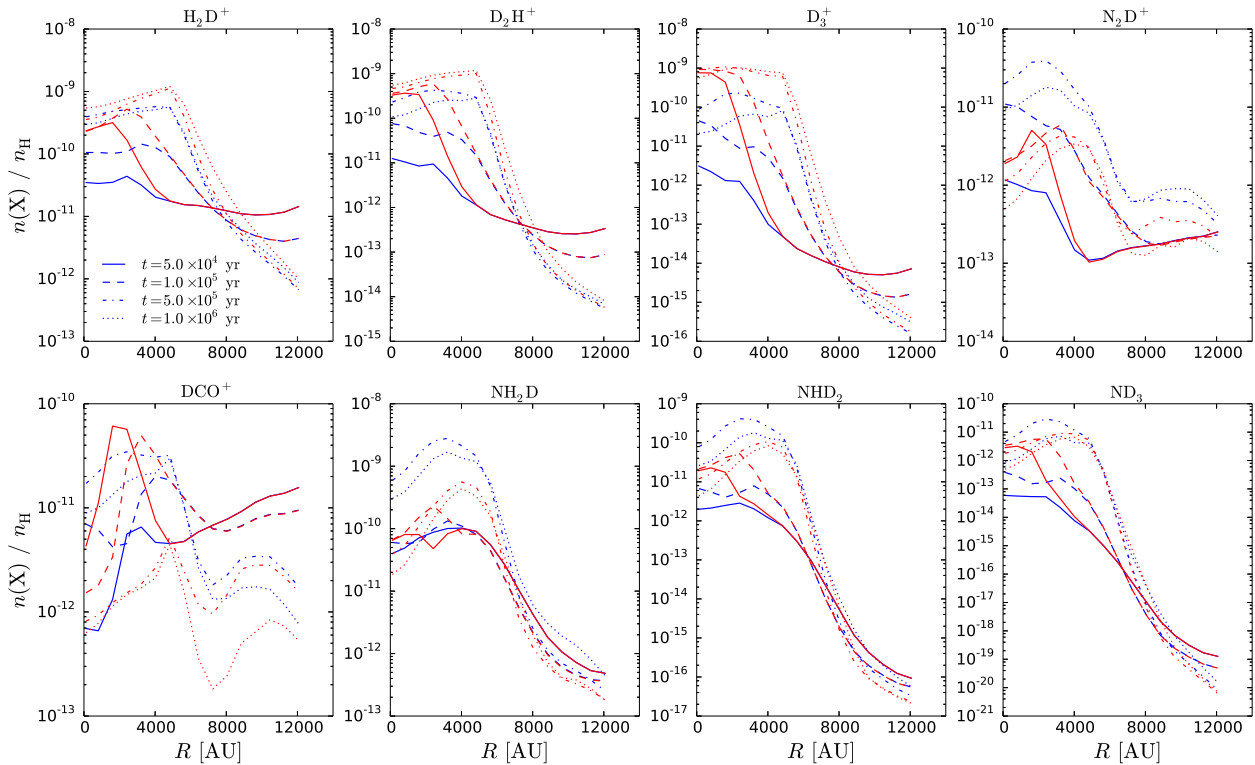}
\caption{Abundances of various deuterated species as functions of radial distance from the core center at different times (indicated in the top left panel). The adopted core model is DT1. The blue lines represent the bulk ice model, while the red lines represent the multilayer ice model.
}
\label{fig:DT1}
\end{figure*}

\begin{figure*}
\centering
\includegraphics[width=1.9\columnwidth]{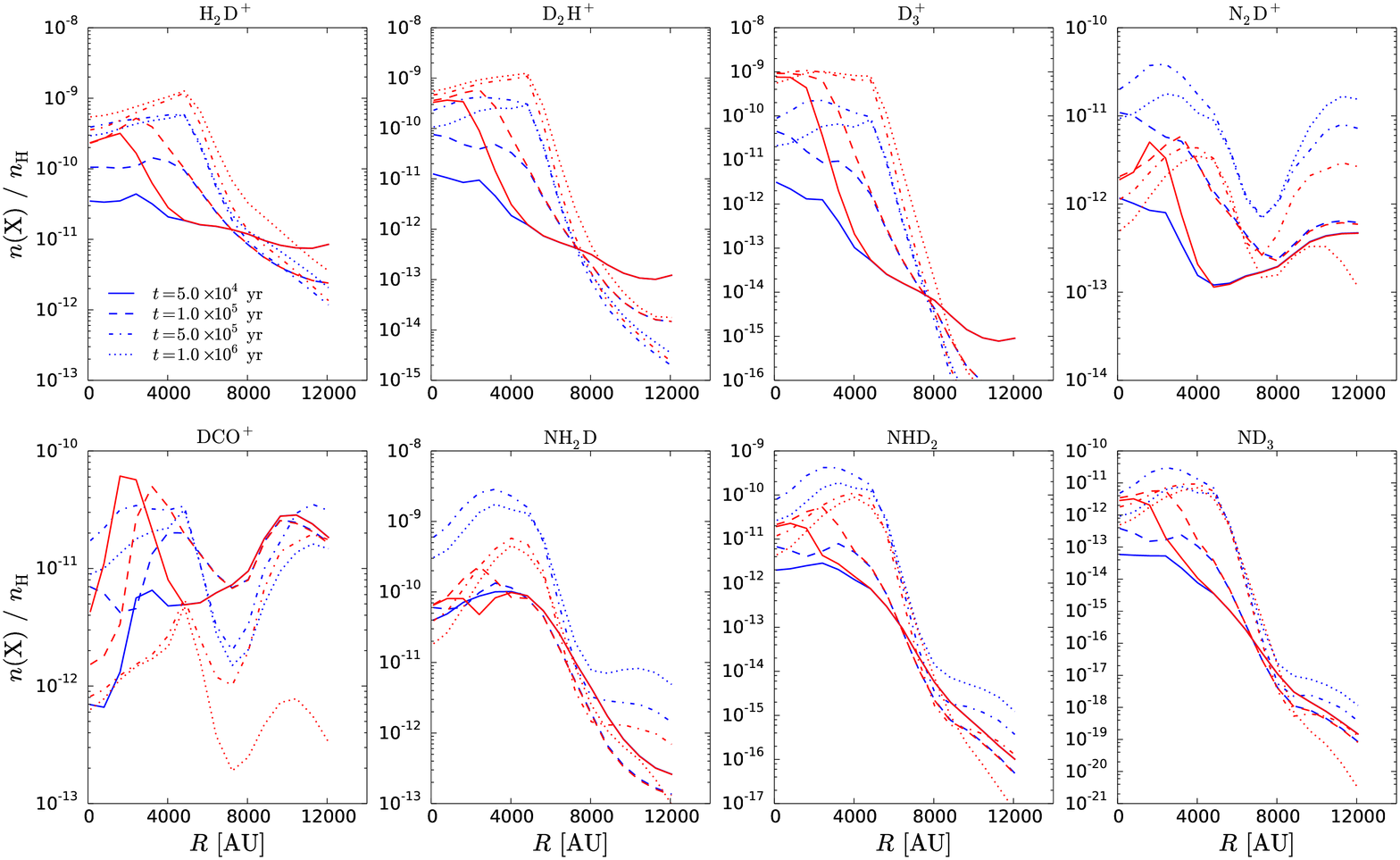}
\caption{As for Fig.\,\ref{fig:DT1}, but adopting the core model DT2.
}
\label{fig:DT2}
\end{figure*}

Figure~\ref{fig:DT1} presents the abundance profiles for a variety of deuterated species at different times in bulk and multilayer models, using the DT1 core (assumed to remain static during the chemical evolution). As expected, the abundances of all of the plotted species are lower near the core edge, where the density is low, than in the high-density environment at the core center. The abundance curves present the same trends already seen in Figs.~\ref{fig:bulk_vs_multi} to \ref{fig:bulk_vs_multi_temp_nh1e6}: in the multilayer model, $\rm H_3^+$ deuteration is enhanced at high density and for deuterated ammonia, the difference between the bulk and multilayer models decreases with the number of D atoms in ammonia. Very low levels of $\rm N_2D^+$ and $\rm DCO^+$ are found in the multilayer model at long timescales, but up to $t \sim 10^5$\,yr the abundance profiles predicted by the two models are somewhat similar to each other. According to the multilayer model, $\rm D_3^+$ becomes the most abundant (deuterated) ion in the core center already very early into the chemical evolution.

\begin{figure*}
\centering
\includegraphics[width=1.9\columnwidth]{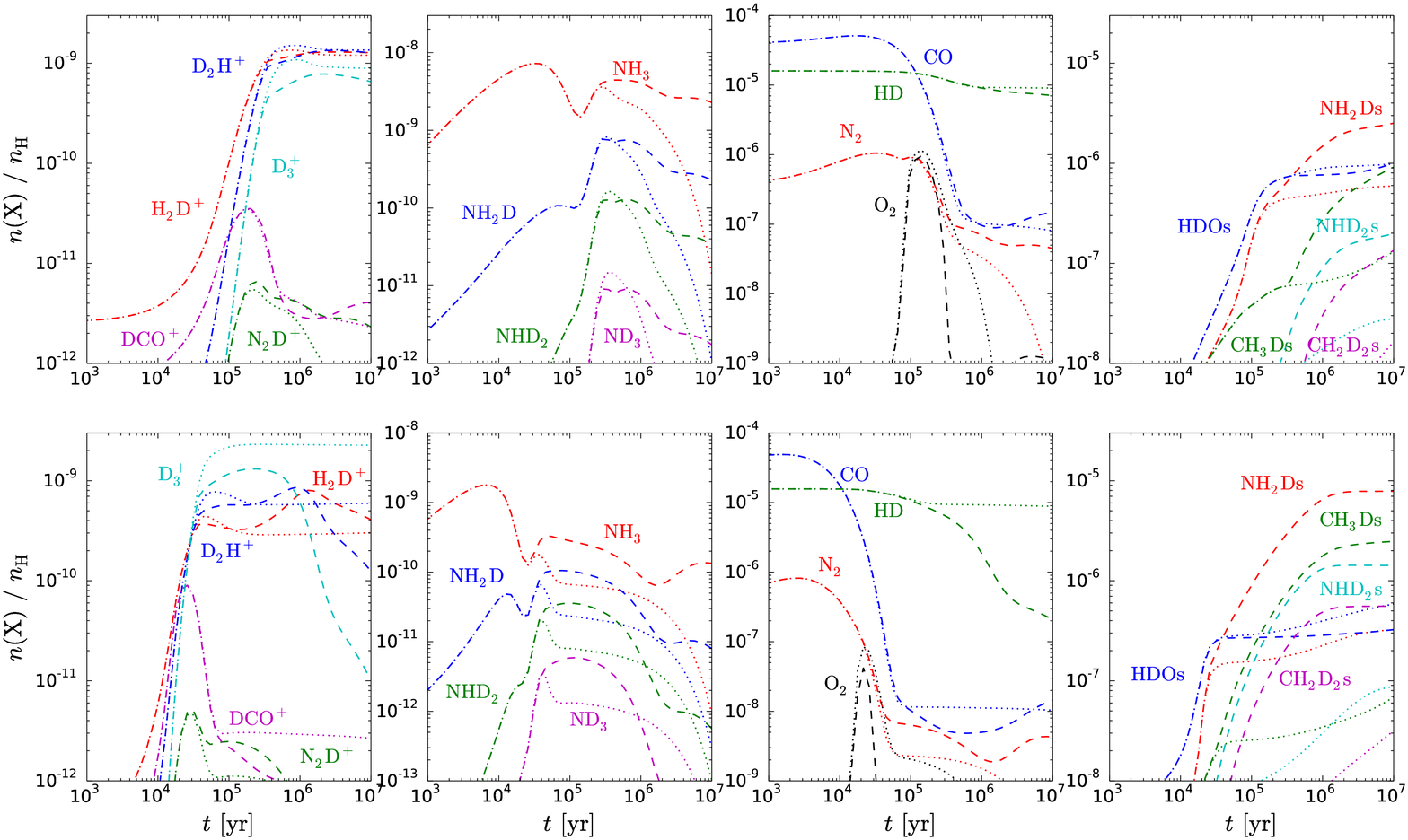}
\caption{Similar to Fig.\,\ref{fig:bulk_vs_multi}, but dashed lines correspond to the multilayer ice model, while dotted lines correspond to the multilayer ice model without mantle desorption.
}
\label{fig:multi_vs_multi_nomantledes}
\end{figure*}

\begin{figure*}
\centering
\includegraphics[width=1.9\columnwidth]{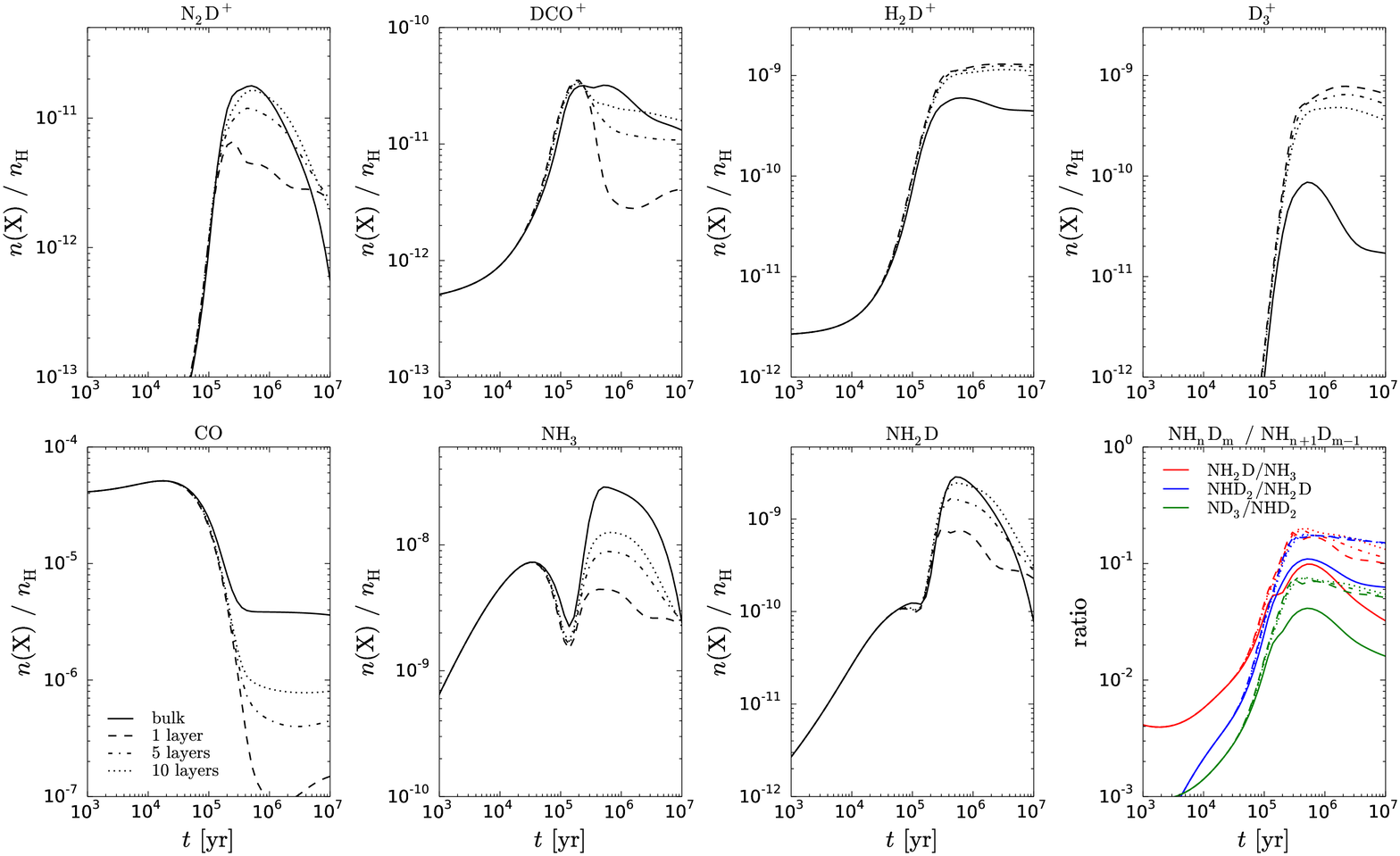}
\caption{Abundances of selected species as functions of time at $n_{\rm H} = 10^5 \rm \, cm^{-3}$ and $T_{\rm dust} = T_{\rm gas} = 10\, \rm K$ in models assuming one or several chemically active ice surface layers (see bottom left panel and the main text for details). Also plotted are the abundance ratios of the various ammonia isotopologs as functions of time (bottom right panel).
}
\label{fig:morelayers}
\end{figure*}

Figure~\ref{fig:DT2} shows the results of calculations otherwise similar to those presented in Figure~\ref{fig:DT1}, but using the DT2 core model. Naturally, no difference is seen between the DT1 and DT2 cases in the innermost parts of the core owing to the equality of the density and temperature profiles (Fig.\,\ref{fig:DT}). Near the core edge, the abundances of $\rm N_2D^+$ and $\rm DCO^+$ (and ammonia) are increased with respect to the DT1 model because of a lower electron abundance and because dissociative recombination reactions with electrons are less efficient at higher temperature, but still very low abundances are found for these species throughout the core. We note that observationally distinguishing between the DT1 and DT2 structures using species like deuterated $\rm H_3^+$ and ammonia which are prominent only in the inner areas would probably be impossible.

\section{Discussion}

\subsection{Multilayer ice model without mantle desorption}\label{ss:nomantledes}

As noted in Sect.\,\ref{ss:modeldesc}, the original \citet{HH93b} multilayer ice model does not include desorption from the mantle, unlike the present work. Even though we limit the mantle desorption to only one ML, it is reasonable to assume that its effect can be non-negligible at least for species that have moderate mantle abundances but relatively low binding energies (for example $\rm N_2$).

Figure~\ref{fig:multi_vs_multi_nomantledes} presents the results of calculations otherwise identical to those shown in Fig.\ref{fig:bulk_vs_multi}, but neglecting mantle desorption, i.e., assuming that the three desorption processes considered here apply to the surface layer only. Evidently, the exclusion of mantle desorption can have a large impact on the abundances at late times. At $n_{\rm H} = 10^5 \, \rm cm^{-3}$, ammonia and $\rm N_2H^+$, and their deuterated forms, suffer particularly much from the exclusion of mantle desorption because of efficient trapping of atomic nitrogen into the mantle, preventing the formation of $\rm N_2$ in the gas. A similar effect is not seen for CO because most of the carbon is locked in CO already at short timescales. It is also evident that HD depletion is less efficient with the exclusion of the mantle desorption, which translates to decreased abundances of the various deuterated species in the ice, and an enhancement of deuterated $\rm H_3^+$, for example. This effect is even more prominent at $n_{\rm H} = 10^6 \, \rm cm^{-3}$. Indeed, at high density and without mantle desorption, the deuterium chemistry resembles the complete depletion scenario studied by \citet{WFP04} and \citet{FPW04}.

\subsection{Multilayer ice model with multiple reactive layers}

It has been shown that H atoms can penetrate into, and react in, several MLs beneath the ice surface \citep{Ioppolo10}, which suggests that reactivity beneath the surface layer should not be neglected. Also, \citet{Vasyunin13} were able to fit the abundances of some complex organics using a (Monte Carlo) model with four active surface layers. Motivated by these studies, we considered extensions of the \citet{HH93b} model where multiple surface layers are chemically active, meaning that all the material in these layers is available for reactions, adsorption and desorption. This is achieved by dividing the surface coverage factor $\alpha$ \citep[Eq.\,(4) in][]{HH93b} by the desired number $n$ of active MLs, which is equivalent to assuming that the surface layer has $n \, \times \, N_{\rm s}$ binding sites (i.e., $n-1$ virtual layers), where $N_{\rm s}$ is the number of binding sites in one ML. The total amount of surface+mantle MLs in the modified multilayer model should of course be nearly identical to a bulk model run in identical physical conditions regardless of the choice of $n$\footnote{But not necessarily exactly identical, because the number of active layers affects the surface chemistry.}, and we have verified through testing that this is the case.

\begin{figure}
\centering
\includegraphics[width=1.0\columnwidth]{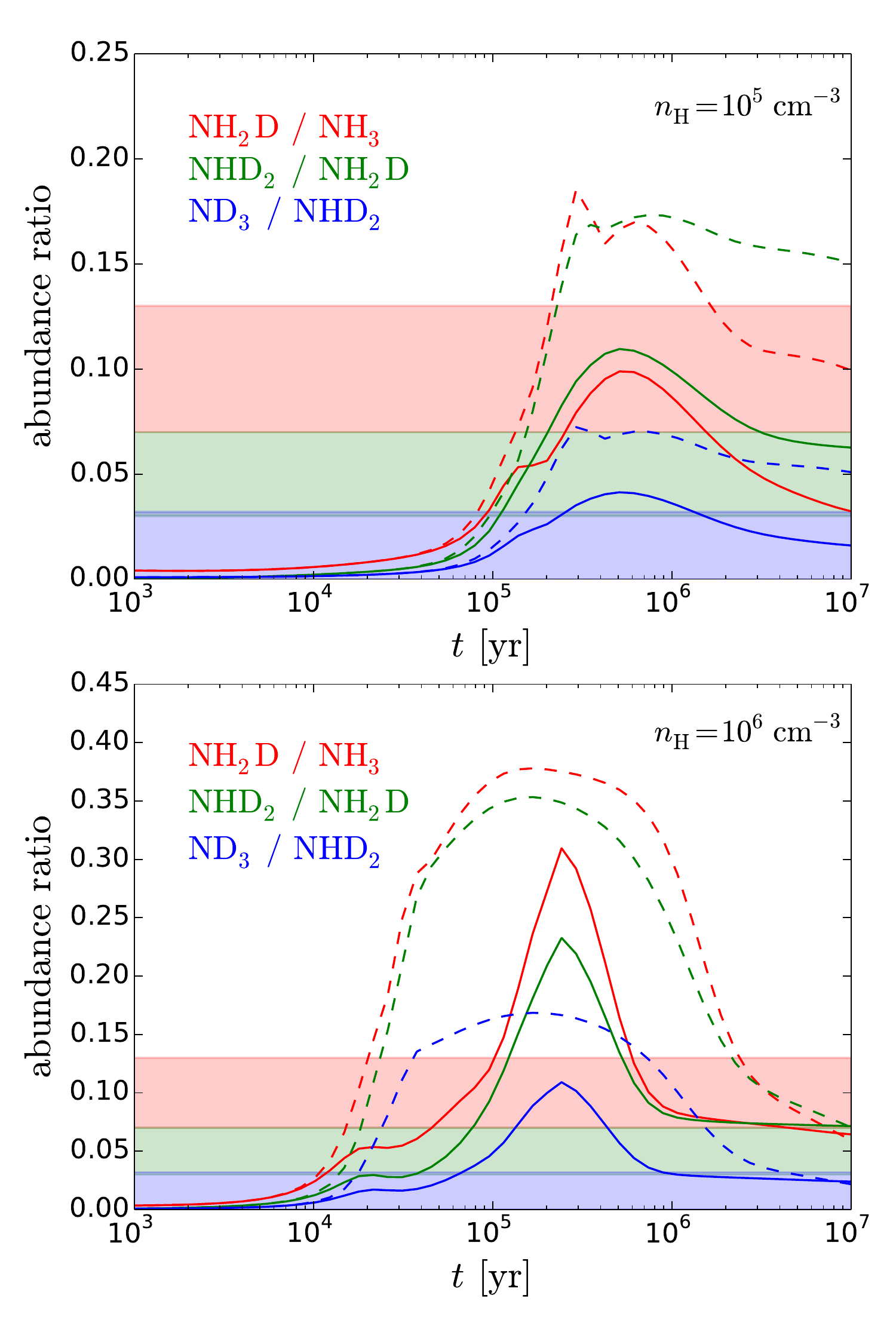}
\caption{Abundance ratios of the ammonia isotopologs, labeled in the figure, as functions of time at two different densities. Solid lines correspond to the bulk ice model, while dashed lines correspond to the multilayer ice model. For reference, the shaded areas show the observed ammonia isotopolog ratios, including errors, toward LDN 134N given in Table~8 of \citet{Roueff05}. For $\rm ND_3 \, / \, NHD_2$, the shaded area represents an upper limit of 0.032.
}
\label{fig:ammonia_ratios}
\end{figure}

Figure~\ref{fig:morelayers} shows the abundances of selected species as functions of time at constant density and temperature in the bulk model (solid lines), the standard multilayer model described earlier in this paper (dashed lines), and two additional models with either 5 or 10 active surface layers (dash-dotted and dotted lines, respectively). Evidently, increasing the amount of active layers brings the $\rm N_2D^+$ and $\rm DCO^+$ abundances closer to the bulk model because their precursors are less depleted. However, the CO abundance is low at long timescales even when 10 MLs are chemically active, and consequently the $\rm H_3^+$ deuteration degree is still high in this model. The ammonia deuteration degree also reaches high values (see Sect.\,\ref{ss:obsimp} for further discussion of this issue). Another test at $n_{\rm H} = 10^6 \rm \, cm^{-3}$ revealed that in these conditions, depletion of CO and $\rm N_2$ is so strong even with 10 active layers that the $\rm N_2D^+$ and $\rm DCO^+$ cannot reach the bulk model values, and the general conclusions for the deuterium fractions in $\rm H_3^+$ and ammonia remain similar to those seen in the lower-density test. The observational implications of the results presented in this as well as the predecing Section are discussed below.

Finally, we note that we cannot recover the exact chemical composition of the bulk model with the modified multilayer model even if we set $n$ equal to the amount of MLs in the bulk model. If there are several tens of active MLs and thus a large amount of accreting material in the surface layer, mantle transfer becomes efficient before the (virtual) surface coverage reaches a value of unity. If $n \sim 100$, the ice morphology reaches a solution at long timescales where the surface coverage is $\sim$$2/3$ and $\sim$$1/3$ of the material is in the mantle. Therefore the method of increasing the number of active MLs discussed here is only reasonable for low values of $n$.

\subsection{Observational implications of multilayer ice chemistry}\label{ss:obsimp}

Observations toward starless/prestellar cores have yielded abundances of the order of $10^{-11} - 10^{-10}$ for $\rm N_2D^+$ \citep{Crapsi05, Vastel06, Miettinen13} and $\rm DCO^+$ \citep{Tafalla06, Schnee07}. Abundances of these orders of magnitude can be relatively easily reproduced by the bulk ice model. However, according to our results, the multilayer ice model cannot reproduce the observed abundances for $\rm N_2D^+$, and for $\rm DCO^+$ high enough abundances are only predicted for early times.

Another constraint on the applicability of the models is given by the deuterium fraction in ammonia, for which the observed fractionation ratios are of the order of $\sim 0.1$ for $\rm NH_2D / NH_3$ and $\rm NHD_2 / NH_2D$, and $\sim 0.01$ for $\rm ND_3 / NHD_2$ (\citealt{Roueff05}; \citealt{Gerin06}). In Fig.\,\ref{fig:ammonia_ratios}, these abundance ratios are shown in the bulk and multilayer models as functions of time at two different densities. Also shown for reference are observed values toward LDN 134N by \citet{Roueff05}. Evidently, the multilayer model produces abundance ratios that are higher than the observed values. We also find that the multilayer model produces a significant enhancement of the $\rm NHD / NH_2$ and $\rm ND_2 / NHD$ ratios (not shown), with multilayer/bulk model peak values of 0.30/0.19 and 0.23/0.12 at $n_{\rm H} = 10^6 \rm \, cm^{-3}$ for $\rm NHD / NH_2$ and $\rm ND_2 / NHD$ respectively. Even if we account for the fact that observations represent averages over the line of sight, it seems clear that the multilayer model would be problematic to reconcile with observations. Temperature effects should not play a large role here, given the similar behavior of the ammonia isotopologs as functions of temperature (Figs.~\ref{fig:bulk_vs_multi_temp_nh1e5} and \ref{fig:bulk_vs_multi_temp_nh1e6}). We note however that definitive conclusions should not be made without proper core modeling, as evidenced by the higher degree of ammonia fractionation observed toward Barnard~1 for example (\citealt{Roueff05}; Table~8).

The multilayer model predicts enhanced deuteration of $\rm H_3^+$. However, the difference in $\rm H_2D^+$ and $\rm D_2H^+$ abundances between the bulk and multilayer models is typically only a factor of a few, as opposed to an order magnitude as found for $\rm N_2D^+$ and $\rm DCO^+$. This suggests that the bulk vs. multilayer scenario could be tested by observing for example the $\rm H_2D^+$ to $\rm N_2D^+$ abundance ratio toward starless cores. According to our results, the ratio is expected to be higher than 100 in the multilayer model, and a few $\times~10$ in the bulk model. For reference, the $\rm H_2D^+$/$\rm N_2D^+$ {\sl column density} ratio derived from observations is $\sim$10 \citep{Caselli02, Caselli03, Crapsi05, Caselli08}. $\rm N_2D^+$ seems a better candidate than $\rm DCO^+$ for comparing the bulk and multilayer approaches, because the abundance of the latter depends strongly on time (Figs.\,\ref{fig:DT1} and \ref{fig:DT2}). Another observational signal of the presence of multilayer chemistry would be a confirmed detection of the dominance of $\rm D_3^+$ over $\rm H_2D^+$ and $\rm D_2H^+$ toward the central parts of starless cores. Unfortunately, observations of $\rm D_3^+$ are not easy because of the lack of permanent dipole moment which means that only vibrational transitions can be observed (in absorption), as also discussed in Section~4 of \citeauthor{FPW04}\,(\citeyear{FPW04}; see also their Table~3).

Summarizing the above, the results presented here imply that the multilayer model is in conflict with observations of deuterated molecules toward starless/prestellar cores, while the bulk model can be more easily reconciled with observations. As the exclusion of mantle desorption was found to lead to even more extreme results in terms of gas-phase deuteration (and depletion of heavy neutrals) than the two-layer mechanism discussed elsewhere in this paper, it seems that more efficient desorption than what is included in the multilayer model discussed here is needed to explain observations of deuterated gas-phase species. On the other hand, allowing for more active MLs on the surface, which naturally leads to more efficient desorption, did not yield a significant improvement in terms of explaning the discrepancy between observations and the multilayer models.

Here we did not include the chemical desorption process in which exothermic reactions on the surface may lead to desorption of the reaction product. It is typically assumed in chemical models that the desorption efficiency, i.e., the percentage of molecules desorbed per reaction event, is around 1\% \citep{Garrod07}, although recent work by \citet{Minissale16} shows that the desorption efficiency may be as high as several tens of \% for some reactions. Adopting this process in the present model is not likely to influence our main conclusions, at least as long as the desorption is limited to the surface layer only.

Finally, the present model is missing reactions between mantle species, but it is unclear what the efficiency of mantle reactions (limited by diffusion of the various species in the mantle) is compared to that of surface reactions, and in any case we are still left with the problem of releasing the mantle species into the gas phase. The effect of alternative non-thermal desorption mechanisms, such as chemical explosions due to cosmic rays \citep{Ivlev15}, on the gas-phase chemistry should be investigated.

\subsection{Comparison to \citet{Taquet14}}\label{ss:taq_comparison}

\begin{figure}
\centering
\includegraphics[width=1.0\columnwidth]{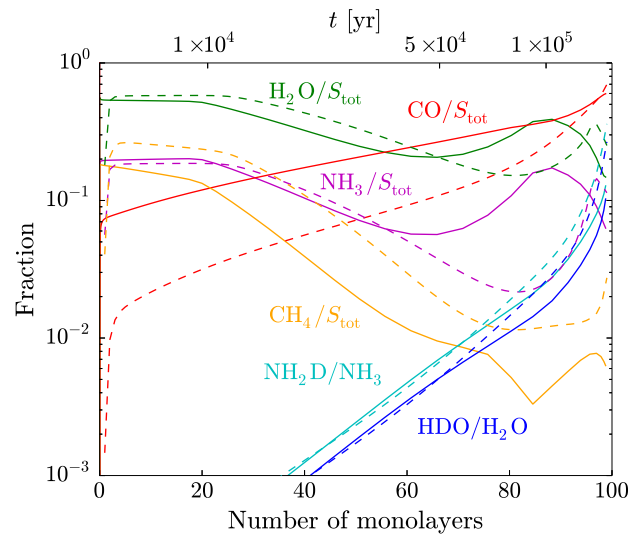}
\caption{Various fractions (see text) on the grain surface calculated with the present chemical model (solid lines) and the \citet{Taquet14} chemical model (dashed lines) as functions of the amount of MLs on the surface. The adopted physical parameters are $n_{\rm H} = 10^5 \, \rm cm^{-3}$, $T_{\rm dust} = T_{\rm gas} = 10\, \rm K$, $A_{\rm V} = 10\, \rm mag$. The time required to form each amount of MLs is indicated on the top axis.
}
\label{fig:surface}
\end{figure}

Recently, \citet{Taquet14} investigated deuterium chemistry in prestellar and protostellar cores using a multilayer chemical model combined with a description of core collapse. Since we use the same model for multilayer chemistry as they do \citep{HH93b}, it is instructive to compare our results with theirs.

We show in Fig.~\ref{fig:surface} the fractional abundances (with respect to the total number density of material on the surface of a grain, represented by $S_{\rm tot}$ in the figure) of four ice species, and two deuterium fractionation ratios, as functions of the amount of MLs on the surface. The plotted curves represent abundances in the reactive surface layer, so that the various fractions represent the composition of the top layer of the ice at the time of its formation. Solid lines correspond to the present model, while dashed lines represent calculations carried out with the GRAINOBLE model from Taquet et al. (\citeyear{Taquet14}; see also Taquet et al. (\citeyear{Taquet12}, \citeyear{Taquet13})) using identical physical parameters and initial chemical abundances, and excluding tunneling. Evidently, the agreement is good, although differences exist between the models as discussed in Sect.\,\ref{ss:chemsets}. At late times in particular the composition of the surface is almost the same in the two models. Notably, the deuterium fraction in water and ammonia is very similar in both models throughout the calculation.

\citet{Taquet14} considered quantum tunneling through activation barriers in surface reactions. Comparing our Fig.~\ref{fig:surface} with Figs.~4~and~5 in \citet{Taquet14} shows that we obtain generally similar ice abundances even though tunneling is not included in the present calculations. The present model does not produce appreciable abundances for species whose formation proceeds through reactions with activation barriers, such as methanol. Still, given the similarity of the modeling results, we qualitatively do not expect the inclusion of tunneling to impact the gas-phase deuteration much since most of the surface deuterium should be locked in deuterated water and ammonia even with tunneling included \citep{Taquet14}. We have verified with test calculations, using both the bulk and multilayer ice models, that the abundances of deuterated gas-phase species indeed remain almost the same even if tunneling is included in the present model.

Finally, we note that the sharp feature seen in the fraction of methane at around 85 MLs (corresponding to a timescale of $\sim 10^5 \, \rm yr$ in this model) is due to a sudden drop in atomic C on the surface, and we verified through testing that the sharpness is smoothed if the time resolution (i.e., the number of timesteps considered) in the model is increased.

\section{Conclusions}

We investigated the abundances of deuterated gas-phase species with a chemical model that includes multilayer ice chemistry adapted from \citet{HH93b}. We included desorption from the ice surface layer and from one monolayer in the mantle. Variations in the density and temperature of the environment were considered, and radial abundance profiles were derived for a core model (a modified Bonnor-Ebert sphere). We compared our results to a more conventional bulk ice approach in which the entirety of the ice (surface+mantle) is available for chemical reactions and desorption, with the aim of studying whether the multilayer model can reproduce the observed abundances of various gas-phase deuterated species as well as the bulk model does.

Our results show that the multilayer model produces very low abundances of $\rm N_2D^+$ and $\rm DCO^+$ at timescales corresponding to starless core lifetimes. The $\rm DCO^+$ abundance can be reconciled with observations in some conditions, but typically only for a very narrow time interval before or after $\sim 10^5$\,yr depending on the density. The $\rm N_2D^+$ abundance is always lower than the observed values. On the other hand, the modeled deuterium fraction in ammonia is much higher than observed. In the absence of mantle desorption, the discrepancy of the models with observed abundances can be even greater, suggesting that desorption from multiple layers in the ice is needed to maintain appreciable abundances of $\rm N_2D^+$ and $\rm DCO^+$ in the gas phase, although tests with an alternative model allowing for multiple chemically active monolayers on the surface did not yield a significant improvement compared to the standard model where only the top layer is active.

Indeed, observations can be more easily reconciled with the bulk ice model. This result is qualitatively puzzling since we expect the ices on grain surfaces to be amorphous, layered structures and it should in principle be challenging to remove material from beneath the few layers closest to the surface, as shown by photodesorption experiments. It is clear that more theoretical and experimental work is needed to learn more about the composition and morphology of interstellar ices, and about the gas-grain chemical interaction, in particular desorption processes. According to our results, the $\rm H_2D^+$ to $\rm N_2D^+$ abundance ratio is higher than 100 in the multilayer model in conditions corresponding to the centers of starless cores, while in the bulk model the ratio is only a few $\times~10$, close to the observed values. Another clear distinguishing feature of the multilayer model is the abundance of $\rm D_3^+$, which becomes the main deuterated ion at high density regardless of the temperatures considered here (6 to 20\,K). A measurement of the abundance of $\rm D_3^+$ compared to $\rm H_2D^+$ and $\rm D_2H^+$ in such environments, perhaps possible through absorption spectroscopy \citep{FPW04}, would also help in constraining the models.

\begin{acknowledgements}

We thank the anonymous referee for helpful comments and suggestions that improved the paper. O.S., P.C., and V.T. acknowledge the financial support of the European Research Council (ERC; project PALs 320620 for O.S. and P.C. and project CHEMPLAN 291141 for V.T.).

\end{acknowledgements}

\bibliographystyle{aa}
\bibliography{multilayer.bib}

\appendix

\section{Models with lower oxygen binding energy}\label{aa:oxygen}

\begin{figure*}
\centering
\includegraphics[width=1.9\columnwidth]{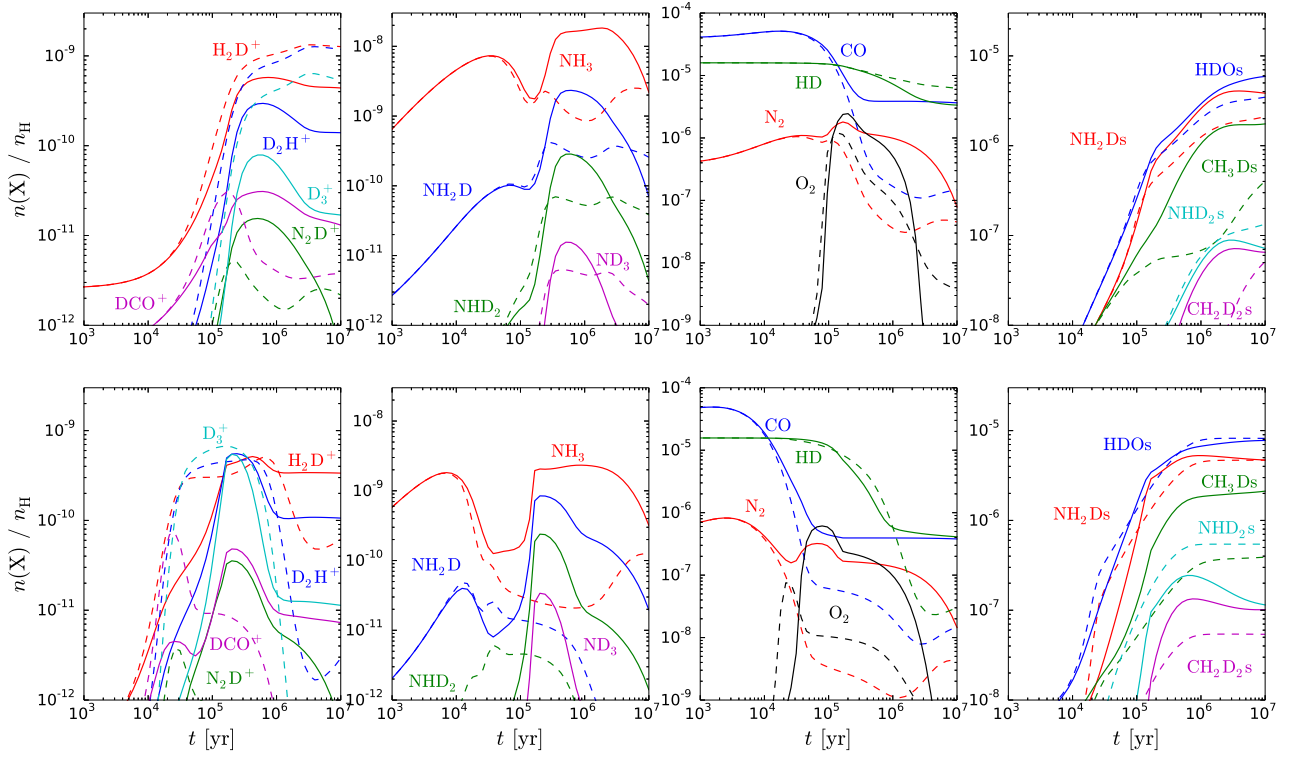}
\caption{As for Fig.\,\ref{fig:bulk_vs_multi}, but calculated assuming oxygen binding energy $E_b(\rm O) = 800$\,K.}
\label{fig:bulk_vs_multi_800}
\end{figure*}

In the results presented in this paper, we adopted an oxygen binding energy of 1390\,K. However, the often-adopted value is $E_b(\rm O) = 800$\,K. For reference, we plot in Fig.\,\ref{fig:bulk_vs_multi_800} the results of calculations otherwise identical to those shown in Fig.\,\ref{fig:bulk_vs_multi}, but assuming $E_b(\rm O) = 800$\,K.

At $n_{\rm H} = 10^5 \rm \, cm^{-3}$, the abundances of the deuterated species in the gas phase are similar to those presented in Fig.\,\ref{fig:bulk_vs_multi}, in both the bulk and the multilayer model. The composition of the ice is more sensitive to the assumed oxygen binding energy, although not by much in the bulk model where the oxygen is free to react with the entirety of the ice. In the multilayer model, decreasing the oxygen binding energy to $E_b(\rm O) = 800$\,K produces much more surface HDO, which becomes the main deuterium carrier in the ice. Notably, $\rm O_2$ is very abundant after $\sim 10^5$\,yr in both the bulk and multilayer models, caused by efficient formation of $\rm O_2$ through $\rm O + OH \longrightarrow O_2 + H$\footnote{This reaction is much more efficient for $E_b(\rm O) = 800$\,K than for $E_b(\rm O) = 1390$\,K, because O depletes less in the former case.}. As noted in Sect.\,\ref{sss:vartemp}, such a high gas-phase $\rm O_2$ abundance is in stark contrast to observations.

At $n_{\rm H} = 10^6 \rm \, cm^{-3}$ in the multilayer model, deuterium is trapped efficiently into HDO ice, and the overall degree of deuteration in the ice is higher than in the $E_b(\rm O) = 1390$\,K case. As a consequence, HD depletes very efficiently from the gas phase, leading to lower abundances of deuterated gas-phase species compared to the models with $E_b(\rm O) = 1390$\,K. However, despite the low abundances, the deuteration degree in ammonia is still very high.

\end{document}